\documentclass[grl]{agutex}

\usepackage{graphicx}
\usepackage{amssymb}
\authorrunninghead{HELMBOLDT}

\titlerunninghead{VLA AND IONOSONDES OBSERVATIONS OF MSTIDS}

\authoraddr{J.\ F.\ Helmboldt, US Naval Research Laboratory, 
Code 7213, 4555 Overlook Ave. SW, Washington, DC 20375 
(joe.helmboldt@nrl.navy.mil)}

\begin{document}

\title{Insights into the nature of northwest-to-southeast aligned ionospheric wavefronts from contemporaneous Very Large Array and ionosondes observations}

\authors{J. F. Helmboldt \altaffilmark{1}}

\altaffiltext{1}{US Naval Research Laboratory, Washington, DC, USA.}

\begin{abstract}
The results of contemporaneous summer nighttime observations of midlatitude medium scale traveling ionospheric disturbances (MSTIDs) with the Very Large Array (VLA) in New Mexico and nearby ionosondes in Texas and Colorado are presented.  Using 132, 20-minute observations, several instances of MSTIDs were detected, all having wavefronts aligned northwest to southeast and mostly propagating toward the southwest, consistent with previous studies of MSTIDs.  However, some were also found to move toward the northeast.  It was found that both classes of MSTIDs were only found when sporadic-$E$ ($E_s$) layers of moderate peak density ($1.5\!<\!\mbox{foEs}\!<\!3$ MHz) were present.  Limited fbEs data from one ionosonde suggests that there was a significant amount of structure with the $E_s$ layers during observations when foEs$>\! 3$ MHz that was not present when $1.5\!<\!\mbox{foEs}\!<\!3$ MHz.  No MSTIDs were observed either before midnight or when the $F$-region height was increasing at a relatively high rate, even when these $E_s$ layers were observed.  Combining this result with AE indices which were relatively high at the time (an average of about 300 nT and maximum of nearly 700 nT), it is inferred that both the lack of MSTIDs and the increase in $F$-region height are due to substorm-induced electric fields.  The northeastward-directed MSTIDs were strongest post-midnight during times when the $F$-region was observed to be collapsing relatively quickly.  This implies that these two occurrences are related and likely both caused by rare shifts in $F$-region neutral wind direction from southwest to northwest.
\end{abstract}

\begin{article}

\section{Introduction}
Traveling ionospheric disturbances (TIDs) are one of the most well known/studied ionospheric phenomena.  So-called medium-scale TIDs, or ``MSTIDs,'' are some of the most commonly observed TIDs with wavelengths and speeds on the order of 100 km and 100 m s$^{-1}$, respectively.  They have been thoroughly studied, mainly with GPS-based data \citep[e.g.,][]{her06,tsu07} and airglow imagers \citep[e.g.,][]{shi03,shi08}, primarily at midlatitudes where they are among the most common impactive disturbances.  They are particularly problematic for high-precision GPS-based applications requiring centimeter-level accuracy \citep{her06}.\par
It has been demonstrated that during nighttime, especially within the summer, the wavefronts of MSTIDs in the northern hemisphere are almost exclusively aligned northwest to southeast, typically propagating toward the southwest \citep{her06,tsu07}.  The origin of this orientation and direction is still not firmly established.  In the past, the orientation of these structures was attributed to the Perkins instability \citet{per73}.  However, the direction of propagation toward the southwest in this case is the opposite of what one would naively expect given the typical eastward-directed background E$\times$B drift.  This was addressed by \citet{kel01} who showed that with a northeastward-directed Pedersen current present resulting from a southeastward-directed neutral wind, the finite extent of the MSTID wavefronts leads to a polarized electric field pointing along the wavefronts toward the northwest.  Thus, E$\times$B drift from this polarized field will push the wavefronts toward the southwest.\par
Despite the qualitative agreement with observations, the \citet{kel01} model suffers from the drawback that it relies on the Perkins instability, which has a growth rate that is too low to generate these features \citep[$\approx \!\!  10^{-4}$ s$^{-1}$;][]{cos04}.  By demonstrating that the Perkins instability is coupled to a separate instability within sporadic-$E$ ($E_s$) layers, \citet{tsu01} and \citet{cos04} showed that northwest-to-southeast aligned features can form within both the $F$-region and the $E_s$ layer with a combined growth rate that is roughly 2--5 times larger.  Since the formation of these structures within the $F$-region is still driven, at least in part, by the Perkins instability, the same polarized electric fields predicted by \citet{kel01} should form along the wave fronts.  For the optimum configuration, \citet{cos04} showed that the horizontal velocity of structures within the $F$-region is proportional to $-E_n \hat{e} + E_e \hat{n}$, implying that the northwestward directed, polarized electric field predicted by \citet{kel01} would cause southwestward motion.\par
While this theoretical framework provides a plausible description of how nighttime, midlatitude MSTIDs are generated, there remains little direct evidence to support it.  In addition, three-dimensional simulations of coupled dynamics between $E_s$ layers and the $F$-region showed that in some cases, polarized electric fields form perpendicular to the northwest-to-southeast aligned wavefronts rather than parallel \citep{yok09}.  In these cases, southwestward motion of the phase fronts was attributed to the southward neutral wind with the coupled $E_s$ layer which was also demonstrated by the simulations of \citet{cos07}.  The \citet{cos07} simulations also showed that the strength of the coupling was reduced when a significant amount of structure was present within the $E_s$ layer.  Recent studies based on observations of disturbances over Japan \citep{shi08} and the southwestern United States \citep{hel12} have also shown that summer nighttime MSTIDs can propagate toward the northeast, sometimes reversing direction from previously being directed southwestward.  These northeastward-directed waves have been tentatively linked to significant drops in $F$-region height, which may provide a clue to the mechanisms that determine the direction of motion of nighttime, midlatitude MSTIDs.\par
To gain further insights into the nature of these disturbances, contemporaneous data from the Very Large Array (VLA) in New Mexico (specifically, $34^{\circ} \: 04' \: 43.497''$ N, $107^{\circ} \: 37' \: 05.819''$ W) and nearby ionosondes have been compiled, spanning 17 nights in June 2002.  Using techniques pioneered by \citet{helint}, VLA observations of multiple cosmic sources were used to characterize the environment of ionospheric fluctuations while using ionosodes data to constrain the properties of both $E_s$ layers and the $F$-region.  The results are presented in \S 2 and discussed further in \S 3.

\section{Fluctuation Spectra, Sporadic-\emph{E}, and \emph{F}-region Compression}
To explore the relationship between MSTIDs and both $E_s$ layers and $F$-region height, a data-set of contemporaneous VLA and ionosondes observations during summer nighttime was compiled.  The VLA data were taken from a larger database generated by the VLA Low-frequency Sky Survey \citep[VLSS;][]{coh07}, a 74 MHz survey of the northern sky conducted between 2001 and 2007.  The VLSS consisted of three or more 20-minute observations for each of 523 overlapping, circular fields, each $15^\circ$ across, yielding a total of 1862 separate observations.  The observations were conducted with the VLA interferometer in its B configuration, an inverted ``Y'' of 27 dishes spanning approximately 11 km.  The correlated signals from pairs of antennas, or ``visibilities,'' were used to image each $15^\circ$ field.  During the imaging process, the affects of the ionosphere across each field were compensated for as a function of time using 1--2 minute ``snapshot'' images of between 20 and 50 relatively bright cosmic sources.  A process referred to as field-based calibration \citep{cot04} uses the position shifts of these sources to map the effects of the ionosphere over the entire field.\par
Each source position shift is proportional to the mean gradient in the total electron content (TEC) over the span of the array along the line of sight to the source.  Given the typical rms noise within a single snapshot, the TEC gradients measured from these position shifts have a precision of about $10^{-4}$ TECU km$^{-1}$ \citep[1 TECU$=10^{16}$ m$^{-2}$;][]{helint}.  Thus, these gradient measurements can be used to probe a much larger range in TEC fluctuation amplitudes than is possible with similar GPS-base data which yield a typical precision on the order of 0.01 TECU \citep[e.g.,][]{her06}.\par
It was shown by \citet{helint} that the mean TEC gradient time series derived from cosmic source position shifts can be used with a straightforward Fourier analysis to produce a three-dimensional spectral cube of TEC gradient fluctuations, one temporal dimension, and two spatial.  This analysis consists of first computing a separate power spectrum cube for each component of the gradient (i.e., north-south and east-west) using temporal and spatial discrete Fourier transforms.  For each spectrum, the impulse response function (IRF) is computed using the positions of the observed sources and the spectrum is normalized by the amplitude of the IRF.  This way, for a simple plane wave, the spectral power will be just the square of its amplitude.  However, since not all observed structures are wavelike, the IRF normalization is denoted in the units reported for the spectra, typically (mTECU km$^{-1}$ IRF$^{-1}$)$^2$, where mTEC is a ``milli-TECU'' (i.e., $10^{-3}$ TECU).\par
Next, the power spectra for the two gradient components are added together so that for a plane wave with spatial frequency $\xi$ and TEC amplitude $A$, the combined spectral power will just be $(2\pi \xi A)^2$.  Note, however, that (1) the spectra are convolved with the IRF which has a full width at half maximum of about 0.009 km$^{-1}$ and (2) many of the observed fluctuations are not wavelike.  Therefore, the combined spectra do not necessarily go to zero at the origin.  For instance, turbulent fluctuations have a power-law spectrum approximately proportional to $\xi^{-11/3}$ for TEC fluctuations and $\xi^{-5/3}$ for the combined TEC gradient spectra \citep[see, e.g.,][]{hel12}.  This spectral analysis has been applied to all 1862 VLSS observations, and a complete climatological investigation based on the resulting spectral cubes is presented by \citet{helvlss}.\par
\begin{figure}
\noindent\includegraphics[width=6in]{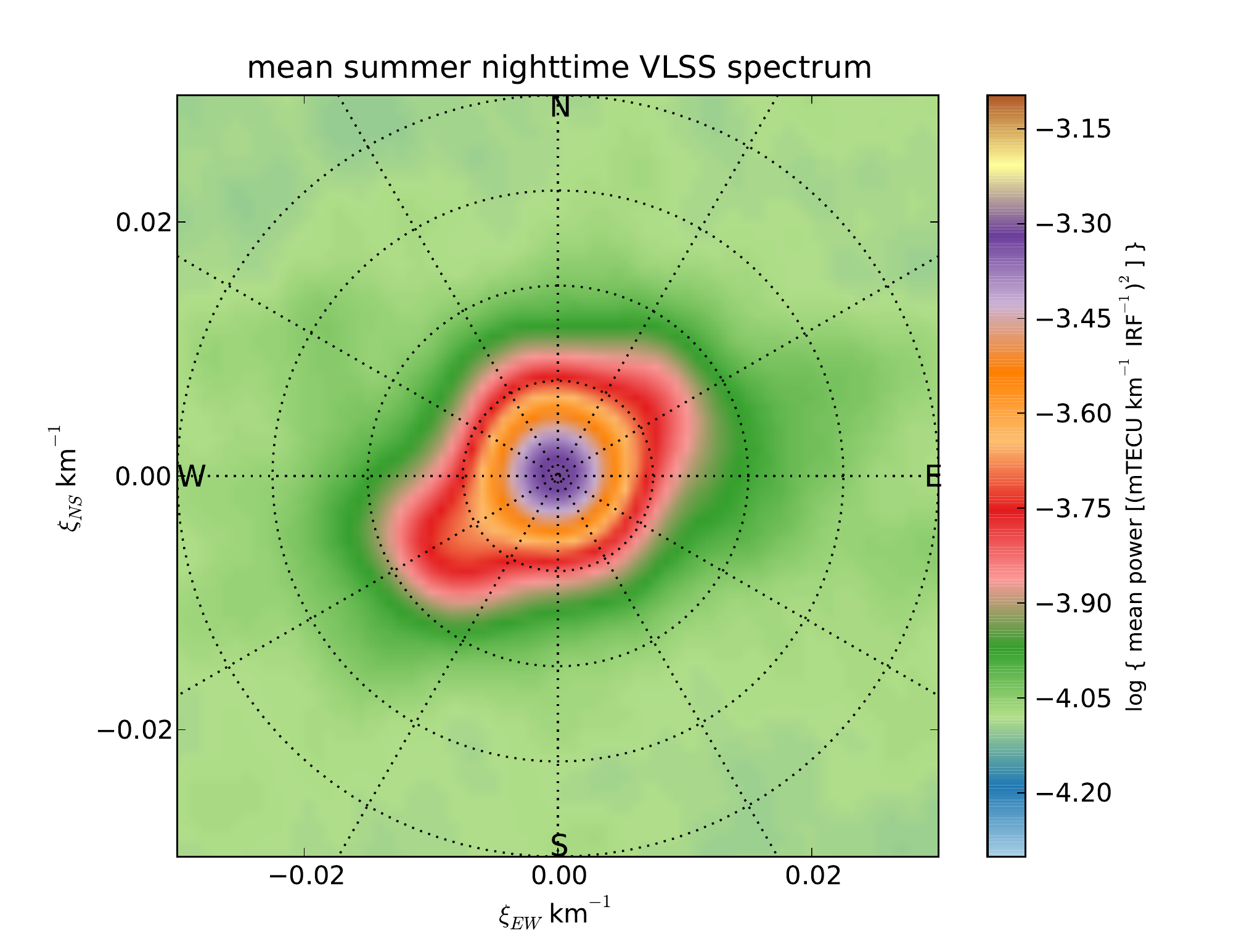}
\caption{The mean TEC gradient fluctuation spectrum over all temporal frequencies for all summer nighttime (20:00--04:00 local time) VLSS observations.}
\label{summer}
\end{figure}

All of the summer observations within the VLSS were conducted in June 2002 (under VLA program number AP441).  The mean nighttime (local times between 20:00 and 04:00) spectral cube from these observations presented by \citet{helvlss} shows a prominent detection of southwestward-directed MSTIDs, as expected.  For reference, the mean spectral power over all temporal frequencies from this average spectral cube is displayed in Fig.\ \ref{summer} as a function of north-south and east-west spatial frequencies, $\xi_{NS}$ and $\xi_{EW}$.  Since there were 132 summer nighttime observations, this provides an excellent opportunity to investigate how the spectral properties of northwest-to-southeast aligned TEC fluctuation wavefronts depend on other ionospheric properties.  In particular, given the discussion in \S 1, it is interesting to investigate how the spectra change with $E_s$ properties and with the height of the $F$-region.\par
\begin{figure}
\noindent\includegraphics[width=6in]{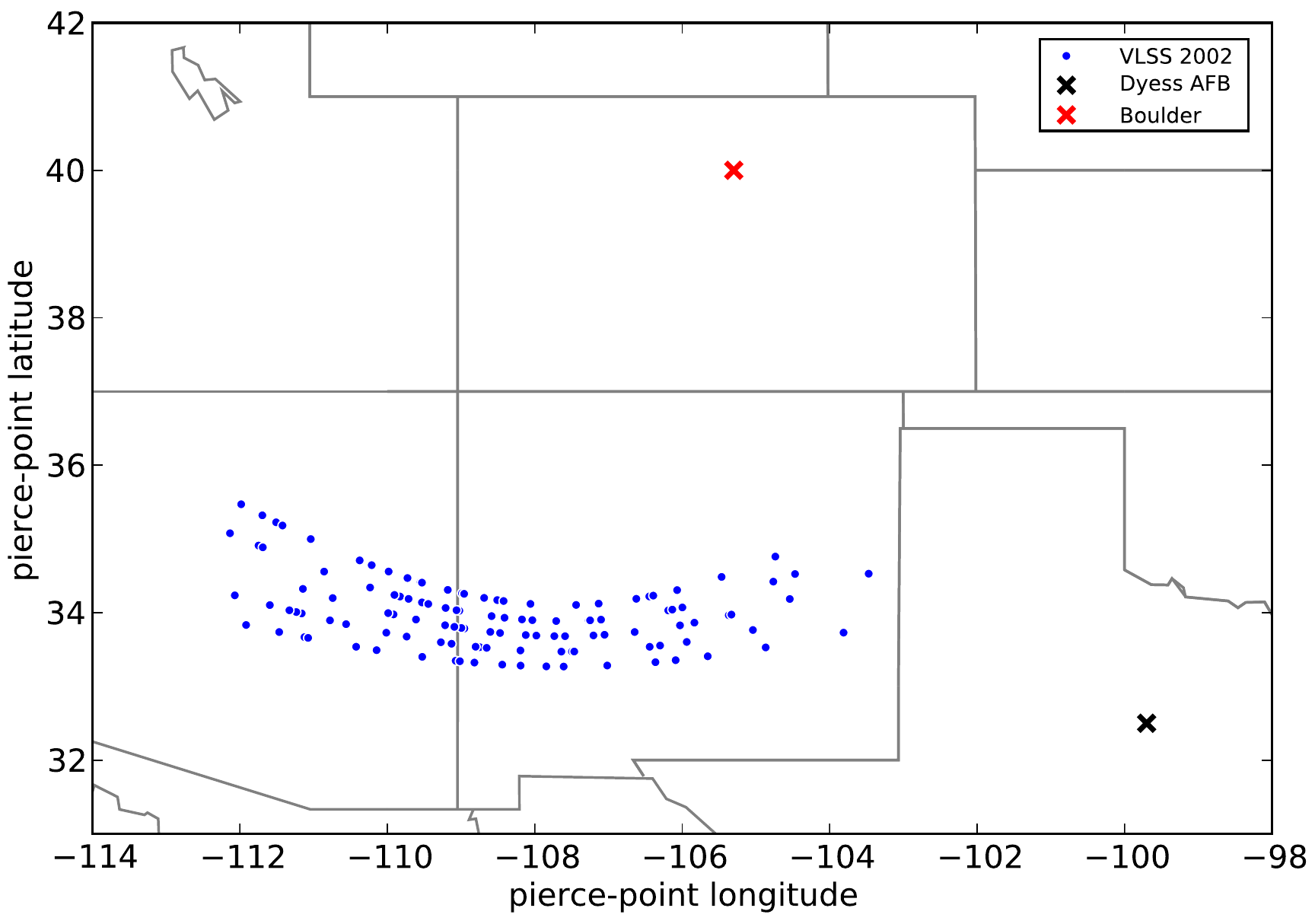}
\caption{The ionospheric pierce points for the VLSS observations conducted during nighttime (20:00--04:00 local time) in June 2002, assuming an ionospheric height of 300 km (blue points).  The locations of the two ionosondes used are plotted as $\times$'s.}
\label{cover}
\end{figure}

To this end, data from nearby ionosondes that were obtained during the June 2002 VLSS observations were compiled.  The two ionosondes closest to the VLSS observations were at Boulder, Colorado and Dyess Air Force Base (AFB) in Texas.  The locations of these stations relative to the ionospheric pierce-points for the 132 summer nighttime VLSS observations, assuming a nominal ionospheric height of 300 km, are shown in Fig.\ \ref{cover}.  One can see that the Boulder station is at a similar longitude, but higher latitude that the VLSS observations; the opposite is true for the Dyess AFB station.  Together the two stations provide a reasonable depiction of ionospheric conditions near the VLSS observations.\par
To constrain the relevant $E_s$ and $F$-region properties, values for the maximum $E_s$ reflection frequency, foEs, and the $F$-region height, $\mbox{h}^\prime\mbox{F}$, were obtained from both stations for each VLSS observation.  Unfortunately, the $E_s$ blanketing frequency, fbEs, was not available from the Dyess AFB station and was flagged as unreliable most of the time (67 out of 132 observations) for the Boulder station during the VLSS observations.  Therefore, the VLSS observations could not be binned according to either fbEs or foES$-$feEs without losing more than half of the observations.  However, the limited Boulder fbEs measurements can be useful for interpreting the results presented below and will be revisited in \S 3.\par
For each VLSS observation, the closest ionosondes observations within $\pm\!20$ minutes were used.  For foEs, 84 out of 132 VLSS observations had measured values from both ionosondes stations, for which the mean foEs between the two stations was used.  The remaining 48 had data from Dyess AFB only.  For $\mbox{h}^\prime\mbox{F}$, several measurements from the Boulder station were flagged as unreliable and were not used, leaving 89 VLSS observations with $\mbox{h}^\prime\mbox{F}$ from Dyess AFB only.\par
During all of the 132 summer nighttime VLSS observations, $E_s$ was detected with at least one of the two ionosondes stations.  The measured foEs from each station ranged from about 1.5--6 MHz.  However, there were several instances where $E_s$ was detected by one station and not the other.  In these cases, foEs$=\!\!0$ was used for non-detections, lowering the mean foEs below 1 MHz.  The summer nighttime VLSS observations were consequently put into four bins based on foEs ranging from 0--6 MHz and a mean fluctuation spectrum cube was computed for each bin.  The mean spectral power over all temporal frequencies is displayed for each bin in the upper row of Fig.\ \ref{coadd} as a function of $\xi_{NS}$ and $\xi_{EW}$.  From these, one can see that both southwestward-directed and northeastward-directed MSTIDs are seen only in the $1.5\!<\!\mbox{foEs}\!<\!3$ MHz bin.  Note that the weaker extension visible in the northeastern quadrant for this foEs bin is not prominent within the mean summer nighttime spectrum shown in Fig.\ \ref{summer}.\par
The results found by \citet{shi08} and \citet{hel12} discussed in \S 1 provide motivation to explore how the direction of propagation of these MSTIDs depends on the change in $F$-region height.  This was done by first estimating the $\mbox{h}^\prime\mbox{F}$ temporal gradient, $\partial\mbox{h}^\prime\mbox{F}/\partial\mbox{t}$, for each VLSS summer nighttime observation.  This was done by numerically computing the temporal derivative using Lagrangian interpolation for each of these observations with at least two other VLSS observations within $\pm40$ minutes.  This excluded only one of the 132 nighttime observations.  It was found that most (54\%) of the observations had $\partial\mbox{h}^\prime\mbox{F}/\partial\mbox{t}$ between $-30$ and 30 km hr$^{-1}$.  Roughly equal fractions had $\partial\mbox{h}^\prime\mbox{F}/\partial\mbox{t}\!<\!-30$ km hr$^{-1}$ (20\%) and $\partial\mbox{h}^\prime\mbox{F}/\partial\mbox{t}\!>\!30$ m hr$^{-1}$ (26\%).\par
The observations with $1.5\!<\!\mbox{foEs}\!<\!3$ MHz were split into three bins, one with $\partial\mbox{h}^\prime\mbox{F}/\partial\mbox{t}\!<\!-30$ km hr$^{-1}$, one with $-30\!<\!\partial\mbox{h}^\prime\mbox{F}/\partial\mbox{t}\!<\!30$ km hr$^{-1}$ and one with $\partial\mbox{h}^\prime\mbox{F}/\partial\mbox{t}\!>\!30$ km hr$^{-1}$.  A map of mean spectral power is displayed in the middle row of Fig.\ \ref{coadd} for each of these bins.  The results are rather striking and show that for $\partial\mbox{h}^\prime\mbox{F}/\partial\mbox{t}\!<\!-30$ km hr$^{-1}$, the northeastward-directed waves are nearly as strong as those directed toward the southwest.  Both classes of MSTIDs are weaker for the $-30\!<\!\partial\mbox{h}^\prime\mbox{F}/\partial\mbox{t}\!<\!30$ km hr$^{-1}$ bin with the southwestward-directed waves as the dominant wavelike feature.  Finally, neither class of waves in present within the spectral map for the $\partial\mbox{h}^\prime\mbox{F}/\partial\mbox{t}\!>\!30$ km hr$^{-1}$ bin.\par
The southwest and northeast features in the $\partial\mbox{h}^\prime\mbox{F}/\partial\mbox{t}\!<\!-30$ km hr$^{-1}$ spectral map seem so symmetric that one may suspect they are the result of an artifact within the Fourier analysis.  However, when examined along the temporal frequency axis, these two features are quite distinct, peaking at two different temporal frequencies.  This is shown in Fig.\ \ref{chan} where maps from the mean spectral cube for temporal frequencies of $\nu=7.5$ hr$^{-1}$ and $\nu=10.5$ hr$^{-1}$ are plotted which represent the peak frequencies for the southwest and northeast features, respectively.  The data used to make the mean spectral cube for this $\partial\mbox{h}^\prime\mbox{F}/\partial\mbox{t}$ bin were also carefully examined and no reason was found to doubt that these features are real detections of separate phenomena.\par
The spectra from observations with $1.5\!<\!\mbox{foEs}\!<\!3$ MHz were also binned by local time into pre-midnight and post-midnight bins.  The mean spectral maps for these two bins are displayed in the bottom panels of Fig.\ \ref{coadd}.  One can see that the pre-midnight spectrum is very similar to the $\partial\mbox{h}^\prime\mbox{F}/\partial\mbox{t}\!>\!30$ km hr$^{-1}$ spectrum in that it shows no signs of wave activity.  Both southwest and northeast features can be seen in the post-midnight map with the southwest feature being the dominant component.  Several combinations of local time bins were tried for the post-midnight observations, but none produced spectral maps with northeastward features as strong as that seen in the map for $\partial\mbox{h}^\prime\mbox{F}/\partial\mbox{t}\!<\!-30$ km hr$^{-1}$.  This indicates that the northeastward-propagating MSTIDs are in fact preferentially seen during times when the $F$-region height is dropping relatively quickly.

\section{Discussion}
\begin{figure}
\noindent\includegraphics[width=6in]{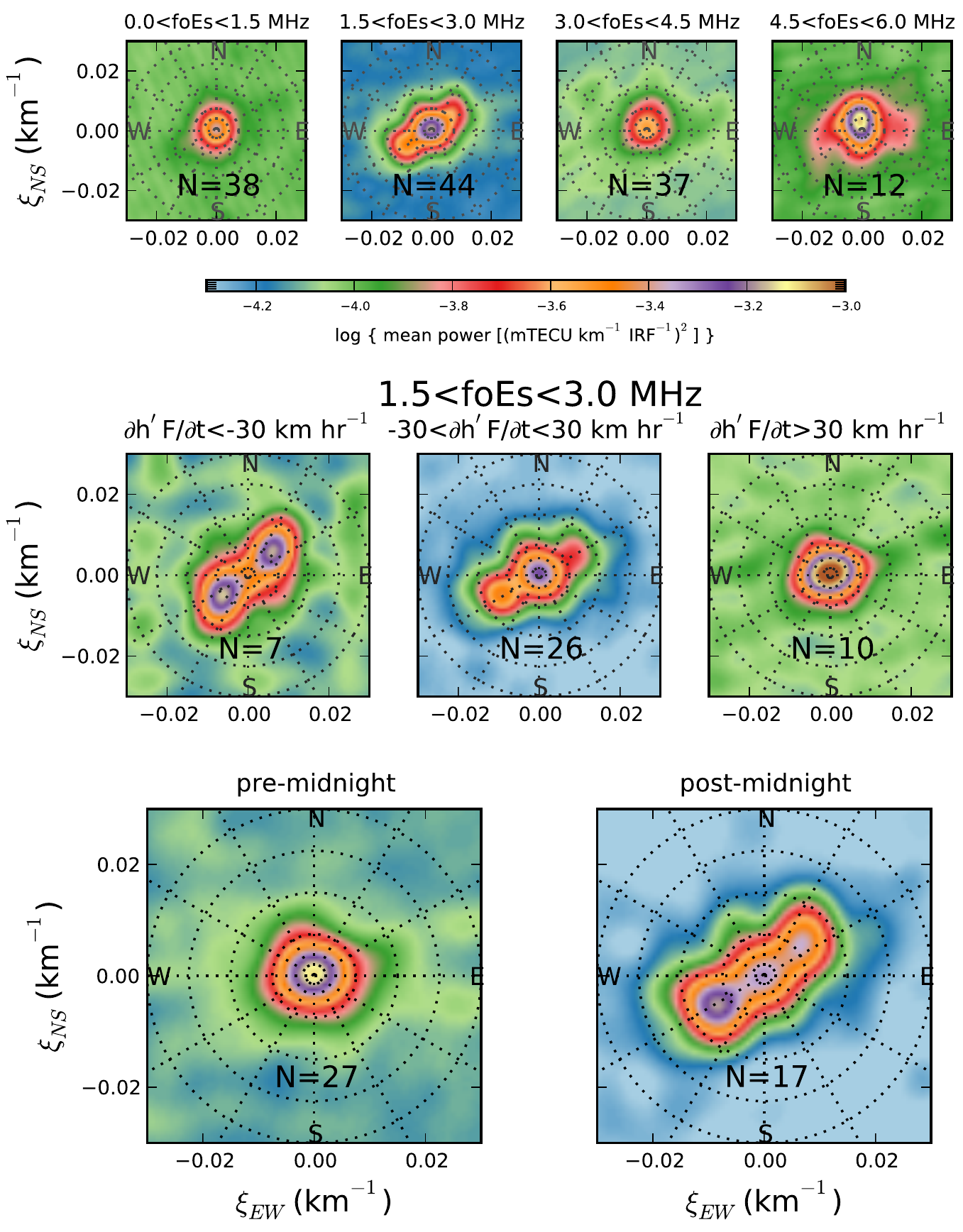}
\caption{Upper:  The mean spectral power (over all temporal frequencies) from observations binned by foEs as a function of north-south and east-west spatial frequency, $\xi_{NS}$ and $\xi_{EW}$.  For observations in the second foEs bin, $1.5\!<\!\mbox{foEs}\!<\!3$ MHz, mean spectral power maps with the observations binned by h$^\prime$F temporal gradient, $\partial\mbox{h}^\prime\mbox{F}/\partial\mbox{t}$, and local time are displayed in the middle and lower rows of panels, respectively.}
\label{coadd}
\end{figure}

\begin{figure}
\noindent\includegraphics[width=6in]{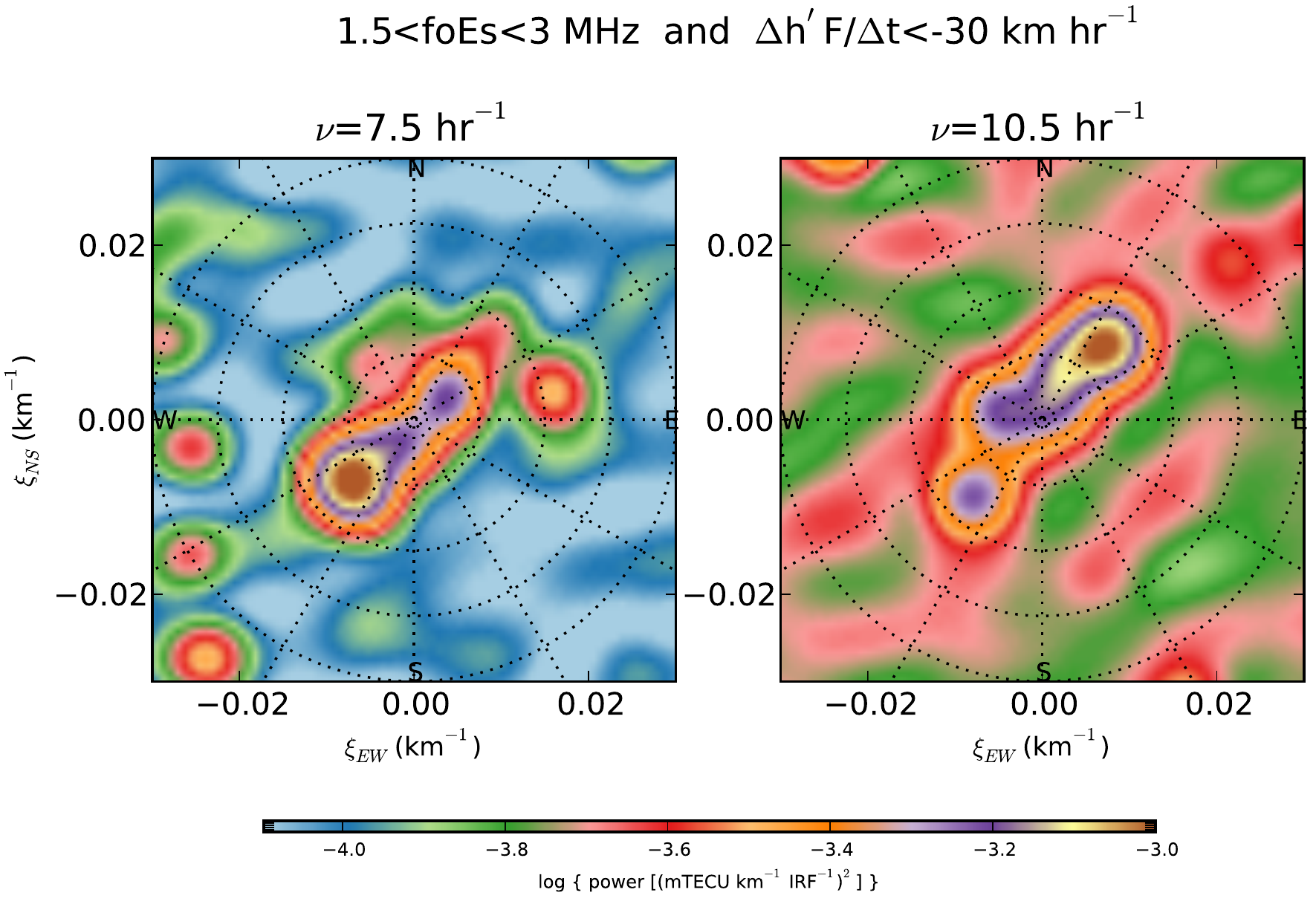}
\caption{For observations where $1.5\!<\!\mbox{foEs}\!<\!3$ MHz and $\partial\mbox{h}^\prime\mbox{F}/\partial\mbox{t}\!<\!-30$ km hr$^{-1}$ (see the lower middle panel of Fig.\ \ref{coadd}), maps of the spectral power at the peak temporal frequencies of the southwestward (left; 7.5 hr$^{-1}$) and northeastward (right; 10.5 hr$^{-1}$) propagating waves.}
\label{chan}
\end{figure}

While not immediately obvious, the results presented in Fig.\ \ref{coadd} are consistent with many aspects of the theoretical framework presented in \S 1 when other ionospheric conditions are taken into account.  First, the fact that summer nighttime MSTIDs were only seen when $E_s$ layers were present is qualitatively consistent with these waves being generated by the $E$-$F$ coupling instability ($EF$CI) described by \citet{cos04}.\par
\begin{figure}
\noindent\includegraphics[width=6in]{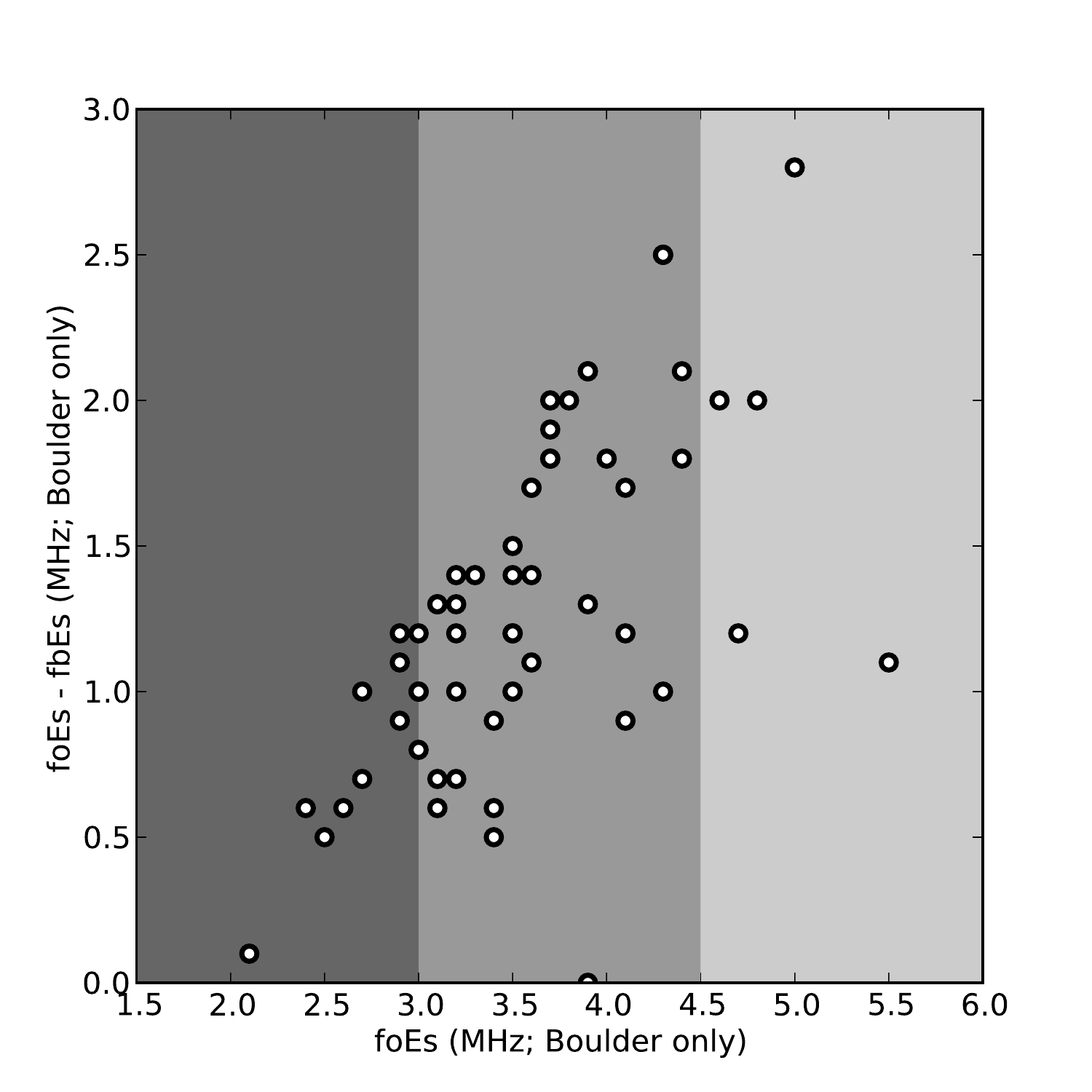}
\caption{For the 65 observations with available measurements for foEs and fbEs from the Boulder station (fbEs was not available from the Dyess station), foEs$-$fbEs versus foEs.  The foEs bins used to make the mean fluctuation spectra shown in the upper panels of Fig.\ \ref{coadd} (excluding the lowest foEs bin) are shaded separately for reference.}
\label{fbes}
\end{figure}

That these waves were preferentially seen when moderate rather than high values of foEs where measured may indicate that $E$-$F$ coupling is generally weaker when there is a larger degree of structure within the $E_s$ layer.  This presupposes that the larger observed values of foEs originate from dense structures within the $E_s$ layers rather than from uniform $E_s$ layers which are simply more dense.  This is supported by the fbEs measurements available from the Boulder station for 65 of the 132 summer nighttime VLSS observations.  Fig.\ \ref{fbes} shows that among the Boulder station data, foEs is near or below the blanketing frequency for observations where foEs$<\! 3$ MHz during which all of the MSTID activity was found.  In contrast, for most of the observations where foEs$>\! 3$ MHz, the difference between foEs and fbEs exceeds 1 MHz.  \citet{oga02} noted that radar echoes from $E$-region structures on scales of tens of kilometers or less [i.e., quasiperiodic (QP) echoes] generally appear when foEs$-$fbEs$>\!1$ MHz.  They also showed that the strength of these QP echoes correlates with foEs$-$fbEs.  Thus, when considered together, the results shown in Fig.\ \ref{coadd} and \ref{fbes} strongly support the notion that an increased level of structure within $E_s$ layers compromises their ability to interact electrodynamicly with the Perkins instability in the $F$-region.  This is consistent with results of the simulations of $E$-$F$ coupling presented by \citet{cos07} that were mentioned in \S 1.\par
Among observations with moderate foEs values, observations that are either in the pre-midnight sector or are associated with larger than normal upward $F$-region motions show no signs of MSTIDs.  This can be explained by taking general conditions during the time of the observations into account.  During the 132 summer nighttime VLA observations used, the mean AE index was 276 nT, indicating that substorm activity was likely common.  In addition, the mean among pre-midnight observations was 315 nT with a maximum of about 690 nT, as apposed to a mean and maximum of 198 nT and 390 nT, respectively, for their post-midnight counterparts.\par
\citet{par71} noted that during substorm activity, zonal electric fields are commonly induced within the ionosphere which typically switch from eastward pre-midnight to westward post-midnight.  The collapse in $F$-region height caused by this shift is often seen near midnight as a negative temporal gradient in TEC within VLA data colloquially referred to as the ``midnight wedge'' \cite[see, e.g.,][Fig.\ 14 and 11, respectively]{kas07,helint}  Thus, it is not surprising that the spectral map for observations with large upward $F$-region motions is similar to those that were conducted before midnight since this is the most likely time for eastward electric fields to exist during substorm activity.  In addition, \citet{cos04} noted that if the background electric field is (magnetic) eastward with little or no north-south component, the $F$-region is stable against the Perkins instability.  Thus, one would expect to see little if any wave activity pre-midnight if the $EF$CI, which depends on the Perkins instability, is chiefly responsible for the growth of northwest-to-southeast aligned wavefronts.  In contrast, the ``resonant'' condition for the $EF$CI given by \citet{cos04} has an electric field with a westward component.  This would imply that wave activity should appear post-midnight when the substorm-induced electric field switches from eastward to westward, which is precisely what was observed.\par
In the post-midnight sector during times when the $F$-region height is relatively stable, the observations show evidence of MSTIDs that are almost exclusively directed toward the southwest.  At midlatitudes in June, \citet{emm03} showed that the neutral winds in the $F$-region are usually directed toward the southwest, especially after midnight.  This implies that it is unlikely that the polarized electric field described by \citet{kel01} contributes significantly to the southwestward motion of these waves since it requires a southeastward-directed neutral wind.  It is more plausible that the background electric field, which was likely westward during the post-midnight VLSS observations, was influencing the motion of these waves as predicted by \citet{cos04}.  However, the typical southward neutral wind velocity within the $E_s$ layers present at the time could also have had a significant effect, similar to what was seen in the simulations of \citet{yok09}.\par
The fact that northeastward-propagating waves were only seen when the $F$-region was dropping at a relatively high rate implies that these drops in height are related to the direction of motion for these waves.  As suggested by \citet{shi08}, the northeastward-directed MSTIDs may be associated with rare instances when the neutral wind shifts direction toward the northwest.  In this case, one would expect polarized electric fields to form along the MSTID wavefronts, similar to that described by \citet{kel01}, but in the opposite direction.  Thus, the motion induced by these fields via E$\times$B drift would be toward the northeast.  In addition, the northward component of the neutral wind would help the westward electric field drive down the $F$-region height at an even higher rate, which matches what was observed.\par
Note that, in order for this scenario to work, the influence of the polarized electric fields along the wavefronts would have to be stronger than those of both the background westward electric field and the southward neutral wind within the $E_s$ layers.  This may only be true during some fraction of the time when northwestward neutral winds are present.  Therefore, even when the $F$-region is collapsing at a higher than normal rate, which may be the signature of a northward meridional wind, the wavefronts could propagate toward the southwest.  The equally strong presence of southwestward-directed waves during times when $\partial\mbox{h}^\prime\mbox{F}/\partial\mbox{t}\!<\!-30$ km hr$^{-1}$ seems to support this.  Three-dimensional simulations similar to those performed by \citet{yok09} with a northwestward neutral wind could potential provide further insights.\par
Finally, note that the results presented here are hampered somewhat by the fact that the VLSS observations were confined to a single month and year during a period when substorm activity was apparently common.  A survey of cosmic sources in the northern sky is being planned that will use the new 330 MHz system being developed for the VLA by the Naval Research Laboratory and the National Radio Astronomy Observatory.  Unlike the VLSS, this survey will be scheduled to optimize time-of-day and time-of-year coverage.  In doing so, it is also likely to sample a wider range in levels of geomagnetic activity than the VLSS observations presented here.  It will also benefit from contemporaneous GPS data from nearly 40 new, continuously operating receiver stations throughout New Mexico that will provide the ability to better characterize medium to large-scale ionospheric structures.  In addition, a new digisonde station will be operating at Kirtland AFB in nearby Albuquerque.  Merging these three data-sets will lead to the development of a comprehensive picture of midlatitude ionospheric disturbances, including $E_s$ and MSTIDs.

\begin{acknowledgments}
Basic research in astronomy at the Naval Research Laboratory is supported by 6.1 base funding.  The VLA is operated by the National Radio Astronomy Observatory which is a facility of the National Science Foundation operated under cooperative agreement by Associated Universities, Inc.  The author is grateful to the UK Solar System Data Centre from which the ionosondes data was obtained and the World Data Center for Geomagnetism, Kyoto for providing AE indices.  The author would also like to thank the reviewers for useful suggestions which greatly improved the paper.
\end{acknowledgments}

\end{article}

\clearpage

\clearpage

\clearpage

\clearpage

\clearpage

\end{document}